\documentclass[twocolumn]{aastex631}
\usepackage{epstopdf}
\usepackage{ae,aecompl}

\usepackage{amsmath}	
\usepackage{amssymb}	
\usepackage{multirow}

\DeclareSymbolFont{matha}{OML}{txmi}{m}{it}
\DeclareMathSymbol{\varv}{\mathord}{matha}{118}

\newcommand{\asec}{$^{\prime\prime}$}
\shorttitle{D/H Galactic Center}
\shortauthors{Colzi et al.}
\graphicspath{{./}{figures/}}

\begin{document}

\title{Deuterium fractionation as a multi-phase component tracer in the Galactic Centre}

\correspondingauthor{Laura Colzi}
\email{lcolzi@cab.inta-csic.es}

\author[0000-0001-8064-6394]{Laura Colzi}
\affiliation{Centro de Astrobiolog\'ia (CSIC-INTA), Ctra. de Ajalvir Km. 4, 28850, Torrejón de Ardoz, Madrid, Spain}
\affiliation{INAF-Osservatorio Astrofisico di Arcetri, Largo E. Fermi 5, I-50125, Florence, Italy}

\author[0000-0003-4561-3508]{Jes\'us Mart\'in-Pintado}
\affiliation{Centro de Astrobiolog\'ia (CSIC-INTA), Ctra. de Ajalvir Km. 4, 28850, Torrejón de Ardoz, Madrid, Spain}

\author[0000-0002-2887-5859]{V\'ictor M. Rivilla}
\affiliation{Centro de Astrobiolog\'ia (CSIC-INTA), Ctra. de Ajalvir Km. 4, 28850, Torrejón de Ardoz, Madrid, Spain}
\affiliation{INAF-Osservatorio Astrofisico di Arcetri, Largo E. Fermi 5, I-50125, Florence, Italy}

\author[0000-0003-4493-8714]{Izaskun Jim\'enez-Serra}
\affiliation{Centro de Astrobiolog\'ia (CSIC-INTA), Ctra. de Ajalvir Km. 4, 28850, Torrejón de Ardoz, Madrid, Spain}

\author{Shaoshan Zeng}
\affiliation{Star and Planet Formation Laboratory, Cluster for Pioneering Research, RIKEN, 2-1 Hirosawa, Wako, Saitama, 351-0198, Japan}

\author[0000-0002-9785-703X]{Lucas F. Rodr\'iguez-Almeida}
\affiliation{Centro de Astrobiolog\'ia (CSIC-INTA), Ctra. de Ajalvir Km. 4, 28850, Torrejón de Ardoz, Madrid, Spain}

\author[0000-0002-5351-3497]{Fernando Rico-Villas}
\affiliation{Centro de Astrobiolog\'ia (CSIC-INTA), Ctra. de Ajalvir Km. 4, 28850, Torrejón de Ardoz, Madrid, Spain}

\author[0000-0001-9281-2919]{Sergio Mart\'in}
\affiliation{European Southern Observatory, Alonso de Córdova, 3107, Vitacura, Santiago 763-0355, Chile}
\affiliation{Joint ALMA Observatory, Alonso de Córdova, 3107, Vitacura, Santiago 763-0355, Chile}

\author{Miguel A. Requena-Torres}
\affiliation{University of Maryland, College Park, MD 20742-2421, USA}
\affiliation{Department of Physics, Astronomy and Geosciences, Towson University, MD 21252, USA}



\begin{abstract}

The Central Molecular Zone (CMZ) contains most of the mass of our Galaxy but its star formation rate is one order of magnitude lower than in the Galactic disc. This is likely related to the fact that the bulk of the gas in the CMZ is in a warm ($>$100 K) and turbulent phase with little material in the pre-stellar phase. We present in this Letter observations of deuterium fractionation (D/H ratios) of HCN, HNC, HCO$^{+}$, and N$_{2}$H$^{+}$ towards the CMZ molecular cloud G+0.693-0.027. These observations clearly show, for the first time, the presence of a colder, denser, and less turbulent narrow component, with a line width of $\sim$9 km s$^{-1}$, in addition to the warm, less dense and turbulent broad component with a line width of $\sim$20 km s$^{-1}$.
The very low D/H ratio $\le$6$\times$10$^{-5}$ for HCO$^{+}$ and N$_{2}$H$^{+}$, close to the cosmic value ($\sim$2.5$\times$10$^{-5}$), and the high D/H ratios $>$4$\times$10$^{-4}$ for HCN and HNC derived for the broad component, confirm the presence of high-temperatures deuteration routes for nitriles. For the narrow component we have derived D/H ratios $>$10$^{-4}$ and excitation temperatures of $7$ K for all molecules, suggesting kinetic temperatures $\le$30 K and H$_2$ densities $\ge$5$\times$10$^{4}$ cm$^{-3}$, at least one order of magnitude larger than for the broad component. The method presented in this Letter allows to identify clouds on the verge of star formation, i.e. under pre-stellar conditions, towards the CMZ. This method can also be used for the identification of such clouds in external galaxies.
\end{abstract}


\keywords{Interstellar molecules - Isotopic abundances - Galactic Center - Star formation}


\section{Introduction} 
\label{sec:intro}

The inner 500 pc of our Galaxy, known as the Central Molecular Zone (CMZ), contains $\sim$80$\%$ of the dense ($>$10$^4$ cm$^{-3}$) molecular gas in the Galaxy ($M_{\rm CMZ}$=2$-$6$\times$10$^7$ M$_{\odot}$; \citealt{morris1996}). However, despite this large reservoir of matter, the star formation rate in the CMZ is at least one order of magnitude lower than in the Disc (\citealt{longmore2013,barnes2017}).

Even if this harsh environment is able to inhibit star formation, it does not stop it completely, and some star formation activity is taking place in the CMZ, as seen from the Arches and Quintuplet clusters, and the younger protoclusters in the Sgr B2 region. The reasons for the suppression of star formation in the CMZ are still not fully understood but are related with its extreme environmental conditions that  provide additional support against gravitational collapse (e.g. \citealt{morris1996}). In particular, the high level of turbulence due to large internal cloud velocity dispersion ($\sim$15--50 km s$^{-1}$, \citealt{morris1996}) and widespread high kinetic temperatures ($T$ from $\sim$50 K to $>$100 K, \citealt{guesten1985}; \citealt{huettemeister1993}; \citealt{ginsburg2016}; \citealt{krieger2017}) could prevent star formation.
In order to understand how star formation proceeds in the CMZ, it is crucial to identify and study the earlier evolutionary stages of the dense, cold and quiescent pre-stellar cores. 
In principle, one expects that for pre-stellar cores to be formed, enough energy (turbulence) has to be dissipated so that gravitational collapse can proceed. Therefore, it is foreseen that pre-stellar cores in the CMZ also present narrower line emission than its surrounding turbulent gas. 
So far, only one progenitor of protocluster in the CMZ have been proposed, the G0.253+0.016 molecular cloud, aka "the brick", which already shows all signposts of massive star formation (\citealt{longmore2012}). However, it is still not clear whether the lack of detection of starless clouds at very early evolutionary stages is directly related to an observational bias. This is due to the presence of warm layers (or envelopes) with large column densities and high velocity dispersion making difficult to observationally disentangle the signatures of the denser and less turbulent pre-stellar gas.

Observation of deuterated molecules is a powerful tool to follow the history of the cold pre-stellar phase of star formation (Caselli $\&$ Ceccarelli 2012), both in low-mass and high-mass star-forming regions (e.g. \citealt{crapsi2005}; \citealt{caselli2008}; \citealt{emprechtinger2009}; \citealt{fontani2011}). 
The high densities ($n>$10$^{5}$ cm$^{-3}$) and low temperatures ($T\leq$30 K) of pre-stellar phases enhance the gas-phase abundance of H$_{3}^{+}$, since its main destroying agent, CO, depletes onto the surface of dust grains. The reaction of H$_{3}^{+}$ with HD (the main reservoir of D in the ISM) produces H$_{2}$D$^+$, which transfers D to other species. As a result, deuterium fractionation in molecules is known to increase their abundance by several orders of magnitude above the D primordial value, D/H=(2.55$\pm$0.03)$\times$10$^{-5}$ (\citealt{zavarygin2018}).

In this Letter, we illustrate how the deuterium  fractionation of molecules (D/H ratio) traces the pre-stellar component embedded in massive warm and turbulent envelopes in the CMZ. We present observations of deuterated species of HCN, HNC, HCO$^{+}$ and N$_{2}$H$^{+}$ towards the warm and highly turbulent molecular cloud G+0.693$-$0.027 (hereafter G+0.693), located in the Sgr B2 complex. In particular, Sgr B2 consists of three main sources, Sgr B2(North), Sgr B2(Main), and Sgr B2(South), which are positioned along a north–south ridge.  Since Sgr B2(N) is at an earlier stage of star formation than Sgr B2(M) (e.g. \citealt{devicente2000}), it seems the star formation activity occurs sequentially from the center to the north. If this is the case, early star formation activity is also expected towards G+0.693 since it is located $\sim$55 arcsec northeast from Sgr B2(N). However, G+0.693 does not show any signposts of ongoing star formation such as ultracompact HII regions, H$_2$O masers, or dust continuum point sources (\citealt{ginsburg2018}). A recent study of gas morphology and kinematics in this region (\citealt{zeng2020}) suggests this cloud to be affected by a cloud-cloud collision producing the shocks likely responsible for its rich chemistry due the sputtering of molecules from dust grains (e.g. \citealt{martin2008}; \citealt{requena-torres2006}; \citealt{zeng2018,rivilla2019}; \citealt{rivilla2020}; \citealt{jimenez-serra2020,rivilla2021a}). In addition to the turbulent gas component (line widths $\ge$15-30 km s$^{-1}$), with a density of about 10$^{4}$ cm$^{-3}$ and a gas temperature $>$100 K (e.g. \citealt{zeng2018}; \citealt{zeng2020}), usually observed in the CMZ, the observed deuterated molecules pinpoint a new component with denser ($n\ge$5$\times$10$^{4}$~cm$^{-3}$), colder ($T\le$30 K) and less turbulent (line widths $\sim$9 km s$^{-1}$) molecular gas component (pre-stellar) in the CMZ, which might be on the verge of gravitational collapse. 
Deuterated species can thus be used as key tracers to disentangle the multi-phase components in the CMZ, and reveal the location and physical conditions of cores where the next generation of stellar clusters will form.

\begin{figure}
\centering
\includegraphics[width=20pc]{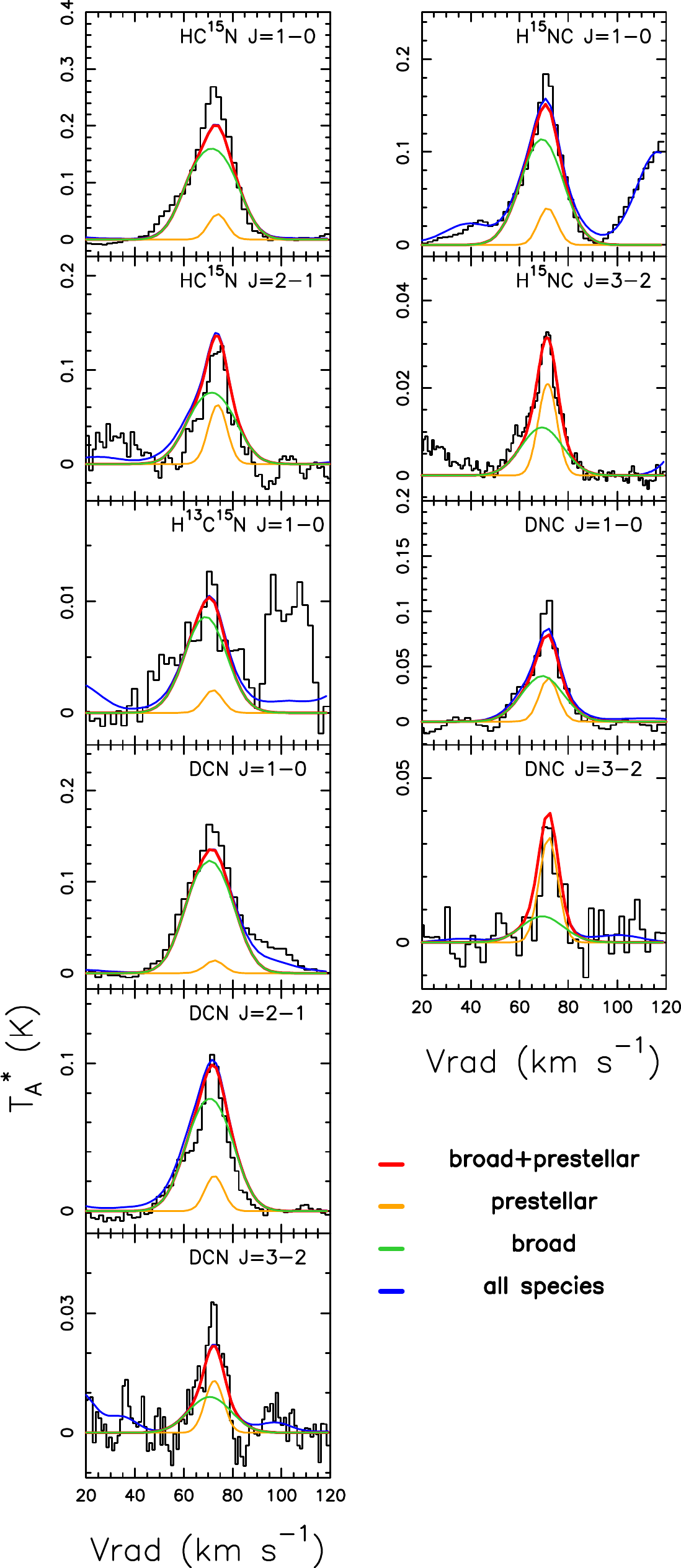}
\caption{Observed transitions of the isotopologues of HCN (left) and HNC (right) studied in this work. The transition shown in each panel are indicated in the upper right. The orange line is the best LTE fit to the pre-stellar component, the green line is the best LTE fit to the broad gas component, and the red line is the sum of the components. The blue line indicates the total modelled line emission, including also the contribution of all molecular species previously identified in the survey (e.g. \citealt{rodriguez-almeida2021a, rodriguez-almeida2021b}; \citealt{rivilla2021a}; \citealt{zeng2021}).}
\label{fig-spectra-hnc}
\end{figure}

\section{Observations}
\label{obs}

The data presented in this work was taken from the high-sensitivity spectral survey towards the G+0.693 molecular cloud (e.g. \citealt{rodriguez-almeida2021a, rodriguez-almeida2021b}; \citealt{rivilla2021a}; \citealt{zeng2021}) carried out  with the IRAM 30m (Granada, Spain) and APEX (Chajnantor, Chile) radiotelescopes . 
The observations, centered at $\alpha_{\rm J2000}$ = 17$^{\rm h}$47$^{\rm m}$22$^{\rm s}$ and $\delta_{\rm J2000}$ = -28$^\circ$21$^{\prime}$27\asec, were made in position-switching mode, with the off position located at (-885\asec,+290\asec) with respect to G+0.693.

The IRAM 30m observations were obtained during 2019, as part of the projects 172--18 (PI Martín-Pintado), 018--19 and 133--19 (PI Rivilla). 
We have used the broad-band Eight MIxer Receiver (EMIR) and the FTS spectrometer (Fast Fourier Transform Spectrometer; \citealt{klein2012}) to cover the frequency ranges 71.76--116.72 GHz, 124.77--175.5 GHz, 199.8$-$222.31 GHz, 223.32--238.29 GHz, 252.52$-$260.30 GHz, and 268.2$-$275.98 GHz, with a frequency resolution of $\sim$800 kHz, corresponding to 0.9-3.3 km s$^{-1}$ at the observed frequencies.
Pointing was checked every 1.5 hours, and focus was corrected at the beginning of the observations and after 4 hours. The HPBW of the telescope varies between 8\farcs9 and 34\farcs3, across the covered frequency range.

The APEX observations were obtained during ten observing runs from July 11 to September 26 2021 in service mode, for a total of 23.4 hr as part of the project 0108.F-9308 (PI Rivilla). 
We used the NFLASH receiver, which allows the simultaneous observations of two  sidebands, each one recorded by two spectrometer processors units (FFTS) of 4 GHz that overlap for 100 MHz, providing a total coverage of 7.9 GHz. 
We observed two different frequency setups centering the upper sideband at 262.0 GHz and 262.3 GHz, respectively. The covered spectral ranges were 243.94--252.12 GHz and 260.18--268.37 GHz, with a spectral resolution of 0.244 MHz ($\sim$0.3 km s$^{-1}$ at 262 GHz). The spectra analysed in this work, which contain H$^{15}$NC(3--2) and H$^{13}$C$^{15}$N(3--2), has been smoothed to $\sim$800 kHz ($\sim$1.1 km s$^{-1}$ at 266 GHz) to match the spectral resolution of the IRAM 30m observations.
The precipitable water vapour (pwv) during the observations was 0.7-3.6 mm.
Focus was performed at the beginning of each observing run, and pointing was checked every 1$-$2 hr. 
The HPBW varies between 23$\farcs$2 and 25$\farcs$6 at the observed frequency range.

The line intensity of the spectra was measured in antenna temperature T$_{\rm A}^{*}$ units since the molecular emission towards G+0.693 is extended over the beam (e.g. \citealt{zeng2020}).

\section{Analysis and results}
\label{res}

\begin{figure}
\centering
\includegraphics[width=20pc]{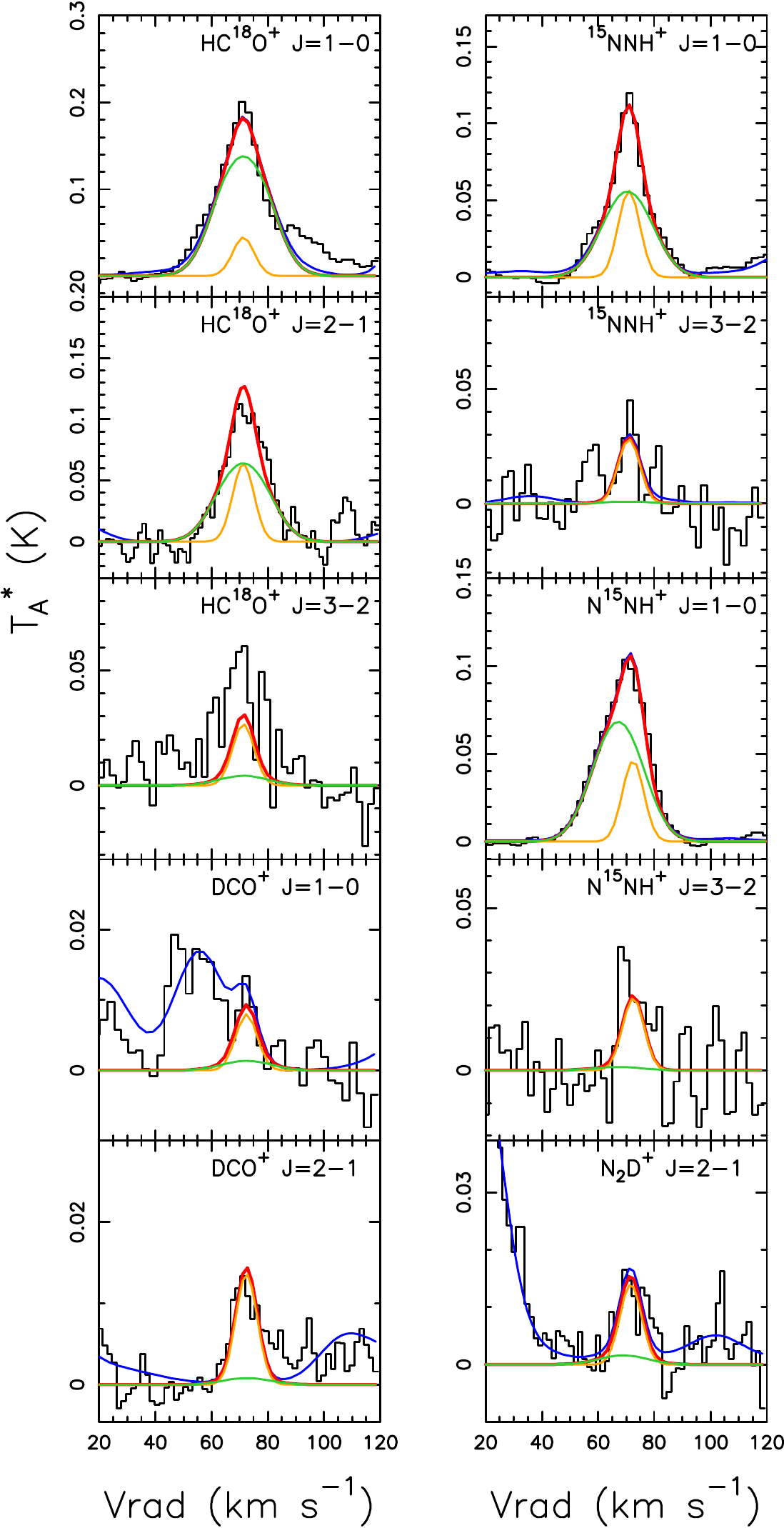}
\caption{Observed transitions of the isotopologues of HCO$^+$ (left) and N$_2$H$^+$ (right) studied in this work. The transition shown in each panel are indicated in the upper right. The orange line is the best LTE fit to the pre-stellar component, the green line is the best LTE fit to the broad component, and the red line is the sum of the two profiles. The blue line indicates the total modelled line emission, including also the contribution of all molecular species previously identified in the survey (e.g. \citealt{rodriguez-almeida2021a, rodriguez-almeida2021b}; \citealt{rivilla2021a}; \citealt{zeng2021}).}
\label{fig-spectra-ions}
\end{figure}

We show the observed spectra of deuterated species of HCN, HNC, HCO$^{+}$ and N$_{2}$H$^{+}$ together with the optically thin isotopologues ($^{13}$C, $^{15}$N, $^{13}$C$^{15}$N, and $^{18}$O) of their hydrogen counterparts in Figs.~\ref{fig-spectra-hnc} and \ref{fig-spectra-ions}. Spectra of transitions blended with other species are shown in Fig.~\ref{fig-spectra-blended}. All the transitions targeted by our survey towards G+0.693 presented in this paper, which include the $J$=1--0, $J$=2--1, and $J$=3--2 transitions, are summarised in Table \ref{transitions}.

Most the observed lines profiles cannot be fitted with a single Gaussian component. The low lying energy transitions ($J$=1--0 and $J$=2--1) clearly show two components with different linewidths (broad and narrow), except for those of DCO$^+$ and N$_2$D$^+$(2--1), and the $J$=3--2 transitions of all molecules that mostly show the narrower component (see Figs.~\ref{fig-spectra-hnc} and \ref{fig-spectra-ions}). As we discuss in Sect.~\ref{disc}, the broad component belongs to the G+0.693 warm and turbulent gas (hereafter "broad" component), usually observed in the emission from transitions of most of molecular species (e.g. \citealt{zeng2018}; \citealt{rivilla2021a}). In contrast, the narrow component observed in the high-\emph{J} lines and for DCO$^{+}$ and N$_{2}$D$^{+}$ reveals for the first time the presence of a less turbulent molecular gas (hereafter "pre-stellar" component). 

We have used the SLIM (Spectral Line Identification and Modeling) tool within the MADCUBA package\footnote{Madrid Data Cube Analysis on ImageJ is a software developed at the Center of Astrobiology (CAB) in Madrid; \url{https://cab.inta-csic.es/madcuba/}.} (\citealt{martin2019})
to identify and perform the multi-line profile fitting.
SLIM generates a synthetic spectrum, assuming local thermodynamic equilibrium (LTE) conditions, and applies an algorithm (AUTOFIT) to find the best non-linear least-squares fit to the data.
The free parameters to be fitted are the column density of the molecule, $N$, the excitation temperature, $T_{\rm ex}$, the peak velocity, $\varv_{\rm LSR}$, and the full-width-half-maximum, $FWHM$ (see details in \citealt{martin2019}). 

To perform the LTE line fitting we have only used the transitions that are not blended with emission from other species (see details in Table \ref{transitions}). The fit procedure requires a total of 8 free parameters to take into account the two velocity components for each molecular species. Thus, to help the fit to converge we have fixed for all the species the $T_{\rm ex}$ and the $FWHM$ to the values as explained below.
To define the best $FWHM$ for the narrow component we have fitted a single Gaussian profile to the transitions for which this component clearly dominates, i.e. $^{15}$NNH$^{+}$ and N$^{15}$NH$^{+}$(3--2), DNC(3--2), N$_{2}$D$^{+}$(2--1), DCO$^{+}$(1--0), and (3--2). The average $FWHM$ derived for these transition is 9$\pm$2 km s$^{-1}$. Then, we have used the same procedure for the broad component using HC$^{18}$O$^{+}$(1--0), DCN(1--0), H$^{15}$NC(1--0), and HC$^{15}$N(1--0), obtaining an average $FWHM$ of 20$\pm$1 km s$^{-1}$.
Thus, we have used for the broad and the pre-stellar components linewidths of 20 km s$^{-1}$ and 9 km s$^{-1}$, respectively, which reproduce well all the observed lines profiles for all molecules (Figs.~\ref{fig-spectra-hnc} and \ref{fig-spectra-ions}).

To derive the $T_{\rm ex}$, we have used  HC$^{18}$O$^{+}$, because it is the only non D-bearing species with three detected transitions ($J$=1--0, 2--1, and 3--2) which appear unblended.
The $T_{\rm ex}$ that better reproduce the three transitions (left panel of Fig.~\ref{fig-spectra-ions}) are 7 K for the pre-stellar component and $\sim$3 K for the broad component. For H$^{15}$NC we find that the broad component will be better fitted with $T_{\rm ex}$ of 4.3$\pm$0.1 K. Given the possible uncertainties in the $T_{\rm ex}$ we used the values of 3 K and 7 K to fit the broad and pre-stellar components in other molecules, except for DNC for which we used the same $T_{\rm ex}$ derived from H$^{15}$NC. The difference in the $T_{\rm ex}$ clearly indicates that the pre-stellar component arises from higher density gas than the broad component with $n\sim$10$^{4}$ cm$^{-3}$  (\citealt{zeng2020}), but it only contributes with 10\% to the total column density. This  will be discussed in Sect.~\ref{disc}.

We applied MADCUBA-AUTOFIT, leaving free $N$ and  $\varv_{\rm LSR}$, with the exception of H$^{13}$C$^{15}$N(3--2), for which the best fit to the line profiles were obtained with $\varv_{\rm LSR}$  fixed to 72 km s$^{-1}$ and to 69 km s$^{-1}$ for the pre-stellar and the broad components, respectively (see Table \ref{table-result}).
The AUTOFIT provides the best solution for the free parameters, and their associated errors. The $\varv_{\rm LSR}$ obtained from the different molecules are consistent with a redshifted  pre-stellar component with respect to the broad component. We are confident that this component is not associated with foreground gas since it would present a different velocity than the bulk of dense gas observed towards the star-forming regions in the Sgr B2 cloud ($\sim$50--70 km s$^{-1}$, e.g. \citealt{goicoechea2002}). We estimate that the total column density can be uncertain by up to 20$\%$ due to the absolute calibration errors.

The derived physical parameters are presented in Table \ref{table-result}, and the best LTE fits to the different line profiles are shown in Figures \ref{fig-spectra-hnc} and \ref{fig-spectra-ions} with coloured solid lines. The orange line corresponds to the pre-stellar component fit, the green line shows the fit to the broad component, and the red line is the sum of the two profiles. The additional blue line also shows the contribution from other molecular species previously identified in the survey (e.g \citealt{zeng2018}; \citealt{rivilla2021a}).

\begin{figure}
\centering
\includegraphics[width=20pc]{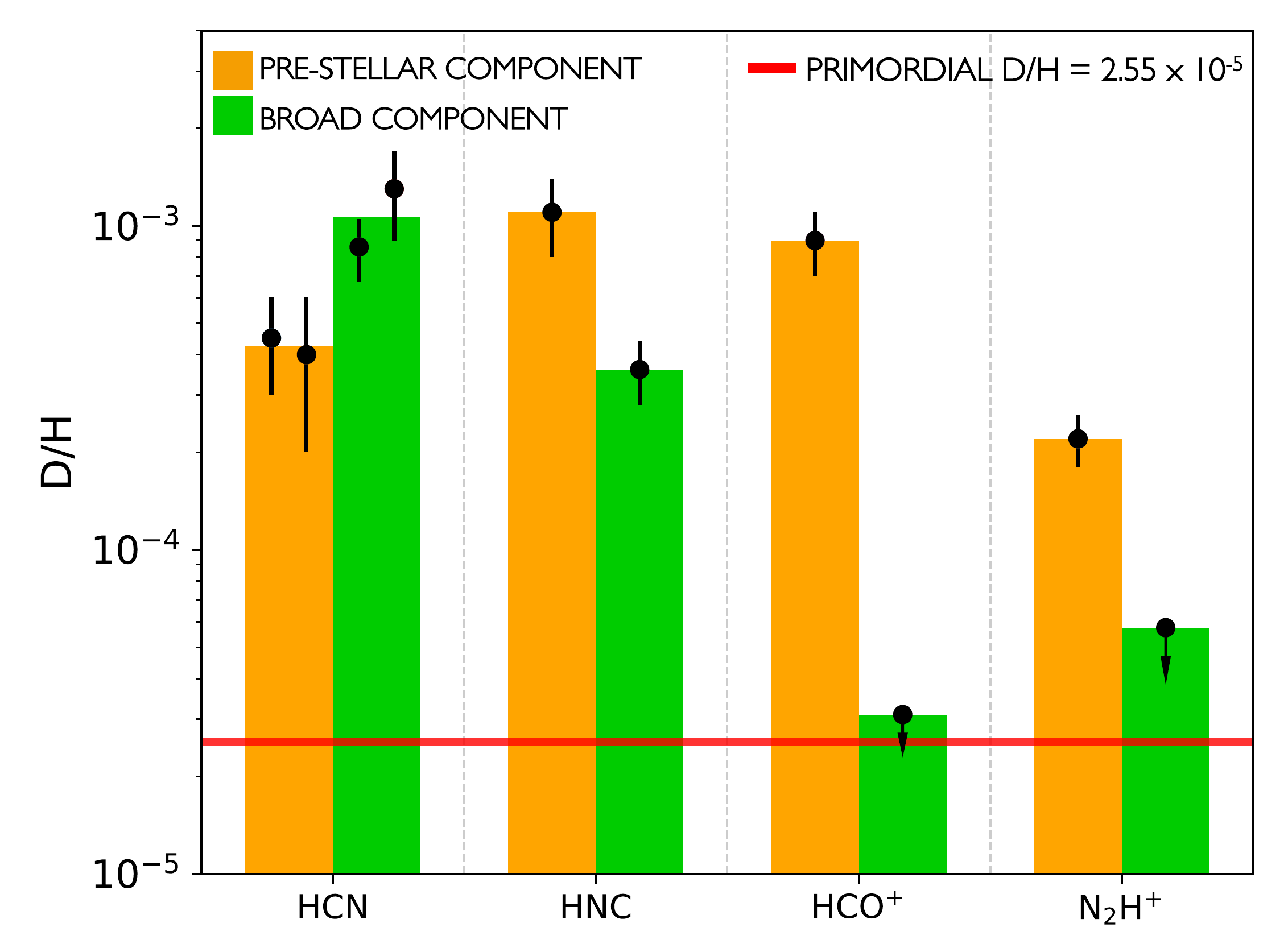}
\caption{D/H ratios found for the different molecules, in the pre-stellar (orange) and broad (green) components. The red horizontal line indicate the primordial D/H value of (2.55$\pm$0.03)$\times$10$^{-5}$ (\citealt{zavarygin2018}).}
\label{fig-histo}
\end{figure}

To compute the D/H ratios, the derived column density of $^{13}$C, $^{15}$N, and $^{18}$O istopologues have been converted to the column densities of the main isotopologues using the typical isotopic ratios found in the CMZ: $^{12}$C/$^{13}$C=20 (\citealt{wilson1994}), $^{14}$N/$^{15}$N=900 (\citealt{guesten1985}), or $^{16}$O/$^{18}$O=250 (\citealt{wilson1994}). The quoted errors in the D/H ratios also consider an additional 20$\%$ error due to the assumed isotopic ratios.
The D/H ratios obtained for HCN and HNC range from $\sim$3.6$\times$10$^{-4}$ up to $\sim$1.3$\times$10$^{-3}$ taking into account both components (see Table \ref{table-result} and Figure \ref{fig-histo}). 
For HCO$^{+}$ and N$_{2}$H$^{+}$, the deuterated species were detected only in the pre-stellar component, giving D/H ratios $>$10$^{-4}$. For the broad component, in contrast, we derived upper limits for the D/H ratios $<$3.1$\times$10$^{-5}$ and $<$5.7$\times$10$^{-5}$ for HCO$^+$ and N$_2$H$^+$, respectively. The upper limits on the total column density of DCO$^{+}$ and N$_{2}$D$^{+}$ have been derived taking into account the 3 $\sigma$ root-mean-square ($rms$) of the integrated intensity at the rest frequencies indicated in Table \ref{table-result}. While the deuteration fraction between the pre-stellar and broad components are $\sim$0.4 and $\sim$3.3 for HCN and HNC, respectively, it increases to $\ge$30 and $\ge$4 for HCO$^{+}$ and N$_{2}$H$^{+}$, respectively. This clearly shows the large differences in the D fractionation for HCO$^{+}$ and N$_{2}$H$^{+}$ with respect to nitriles in both components.

We have also checked if the differences found in the isotopic compositions of the two components are still significant when considering different $FWHM$ and $T_{\rm ex}$ than those used for the LTE fit. To do that, we have explored different values of these parameters for the two components and for which the fit converges. In particular, for the narrow component we have explored $T_{\rm ex}$ between 5 and 15 K, and for the broad component between 3 and 6 K. 
We have found that for the narrow component, the D/H ratios vary by factors of 1-1.8 and 1-3.4 with respect to those shown in Table \ref{table-result} for HNC and N$_{2}$H$^+$, respectively, for the temperature ranges considered.
For the broad component, the D/H ratio of HNC varies by factors 0.9-1.3, and the D/H upper limit of N$_{2}$H$^+$ varies by factors 0.4-1 with respect to those shown in Table \ref{table-result}. 
In the range of $T_{\rm ex}$ used, the deuteration fraction between the narrow and the broad component varies only between 3.3 and 4.3 for HNC, and between $>$4 and $>$19 for N$_{2}$H$^+$. Thus, the differences in the isotopic composition of the two components and between the different species are still significant despite the possible uncertainties in $T_{\rm ex}$.

We have also explored the effects of changes in the $FWHM$ between 9 km s$^{-1}$ and 11 km s$^{-1}$ for the narrow component and between 15 km s$^{-1}$ and 25 km s$^{-1}$ for the broad component. In this case, for the narrow component, the D/H ratios vary by factors of 1-1.45 and 1-1.2 with respect to those shown in Table \ref{table-result} for HNC and N$_{2}$H$^+$, respectively, for the $FWHM$ ranges considered.
For the broad component, the D/H ratio of HNC varies by factors 1-1.14, and the D/H upper limit of N$_{2}$H$^+$ varies by factors 1-1.25 with respect to those shown in Table \ref{table-result}. In the range of possible $FWHM$, the deuteration fraction between the narrow and the broad component varies between 3.3 and 4 for HNC, and between $>$2 and $>$4 for N$_{2}$H$^+$. Thus, we conclude that the differences in the isotopic composition of the two components and for the two species are still significant even assuming rather extremes values of line widths and $T_{\rm ex}$.

\begin{table}
\setlength{\tabcolsep}{2pt}
\begin{center}
\caption{\label{table-result} Results from the LTE fitting procedure.}
\begin{tabular}{lccccc}
\hline
Molecule & $FWHM$	& $\varv_{\rm LSR}$	& $T_{\rm ex}$ & $N$ & D/H 	\\
& (km s$^{-1}$) & (km s$^{-1}$)   & (K) &  ($\times$10$^{12}$ cm$^{-2}$) & ($\times$10$^{-4}$) \\
\hline
\multicolumn{6}{c}{HCN}\\
\hline
DCN & 9 & 72.5$\pm$0.8 & 7 & 0.33$\pm$0.06 & \\
DCN & 20 & 70.7$\pm$0.3 & 3 & 39.8$\pm$1.4 & \\
\hline
HC$^{15}$N &9 &  73.9$\pm$0.6 & 7 & 0.81$\pm$0.15 & 4.5$\pm$1.5\\
HC$^{15}$N &20 &  71.5$\pm$0.6 & 3 & 51$\pm$4 & 8.6$\pm$1.9\\
\hline
H$^{13}$C$^{15}$N & 9 & 72 & 7 & 0.072$\pm$0.019 & 4$\pm$2\\
H$^{13}$C$^{15}$N & 20 & 69 & 3 & 1.4$\pm$0.2 & 13$\pm$4                          \\                                         
\hline
\multicolumn{6}{c}{HNC}\\
\hline
DNC & 9  &  71.9$\pm$0.5  & 7 & 0.83$\pm$0.11& \\ 
DNC& 20  &  69.4$\pm$1.1  & 4.3 & 2.4$\pm$0.3 & \\
\hline
H$^{15}$NC & 9  &  71.55$\pm$0.14 &  7 & 0.84$\pm$0.04 & 11$\pm$3\\
H$^{15}$NC & 20  &  69.22$\pm$0.16 &  4.3$\pm$0.1 & 7.48$\pm$0.16 & 3.6$\pm$0.8\\  
\hline
\multicolumn{6}{c}{HCO$^{+}$}\\
\hline
DCO$^{+}$ & 9  &  72.4$\pm$0.5 &  7 &  0.12$\pm$0.01 & \\ 
DCO$^{+}$ & 20  &  70&  3 &  $\le$0.2\tablenotemark{a} &  \\   
\hline
HC$^{18}$O$^{+}$ & 9  &  71.3$\pm$0.6 &  7 &  0.48$\pm$0.06 & 9$\pm$2\\
HC$^{18}$O$^{+}$ & 20  &   71.3$\pm$0.5 &  3 &  23.4$\pm$1.5 & $\le$0.31\\
\hline
\multicolumn{6}{c}{N$_{2}$H$^{+}$}\\
\hline
N$_{2}$D$^{+}$ & 9  &  71.7$\pm$0.8 & 7 & 0.15$\pm$0.02& \\
N$_{2}$D$^{+}$ & 20  &  69	& 3 & $\le$0.5\tablenotemark{b}&  \\
\hline
$^{15}$NNH$^{+}$ & 9  & 71.01$\pm$0.5 & 7 & 0.74$\pm$0.09 & 2.2$\pm$0.4\\
$^{15}$NNH$^{+}$ & 28  &  70.5$\pm$0.9 & 3 & 8.9$\pm$0.9 & $\le$0.64\\
\hline
N$^{15}$NH$^{+}$ & 9  & 72.5$\pm$0.5 & 7 & 0.59$\pm$0.09 & 2.7$\pm$0.8\\
N$^{15}$NH$^{+}$ & 20  &  67.2$\pm$0.7 & 3 & 11.2$\pm$0.9 & $\le$0.51\\
 \hline
  \normalsize
   \end{tabular}
   \end{center}
\tablecomments{Parameters without errors are fixed in the fitting procedure, as explained in Sect.~\ref{res}.}
\tablenotetext{a}{Upper limit derived taking into account the $rms$ of $\sim$1.1 mK at 144.083 GHz.} \tablenotetext{b}{Upper limit derived taking into account the $rms$ of $\sim$2.7 mK at 154.22 GHz.} 
 \end{table}

\section{Discussion and Conclusions}
\label{disc}

The Sgr B2 cloud complex is one of the most active sites of star formation in our Galaxy. As already explained in Sect.~\ref{sec:intro}, towards its densest part, this molecular cloud harbours three well-known massive star-forming clusters: Sgr B2(N) (in the north), Sgr B2(M) (in the centre), and Sgr B2(S) (in the south) that comprise numerous compact and ultra-compact HII regions (e.g. \citealt{gaume1995}; \citealt{schmiedeke2016}; \citealt{ginsburg2018}). However, the presence of filament-, arc- and shell-shaped dense gas likely produced by stellar feedback (\citealt{martin-pintado1999}), and of a large population of high-mass protostellar cores found by \citet{ginsburg2018}, indicate that star formation is not restricted to the central cloud region, but it is also taking place in the extended envelope of Sgr B2.
Since Sgr B2(M) is considered to be more evolved than Sgr B2(N) and (S) (e.g. \citealt{devicente2000}; \citealt{ginsburg2018}), star-forming activity may proceed sequentially from Sgr B2(M) and outwards, to continue towards Sgr B2(N) in the north and Sgr B2(S) in the south. In the scenario of sequential star formation, it is expected that G+0.693 is in an earlier evolutionary phase since it is located northeast of Sgr B2(N).

Previous observations towards the quiescent clouds in the GC traced only the turbulent warm gas in a very turbulent component unable to form stars. Our observations of deuterated species reveal, for the first time, very low D/H ratios in HCO$^{+}$ and N$_{2}$H$^{+}$, close to the cosmic value, for the turbulent broad component and the presence of a new narrow ($FWHM$ of 9 km s$^{-1}$) quiescent component, with high levels of deuteration in HCN, HNC, HCO$^{+}$ and N$_{2}$H$^{+}$ (D/H $>$ 10$^{-4}$, see Fig.~\ref{fig-histo}). 
This gas is clearly less turbulent (narrower $FWHM$) than the surrounding gas (broad gas component with a $FWHM$ of 20 km s$^{-1}$), exhibiting  similarly properties to those found in Galactic disc pre-stellar cores. The different degrees of D fractionation found in HCO$^{+}$ and N$_{2}$H$^{+}$ for both components indicates that the pre-stellar component should have a kinetic temperature $T\le$ 30 K, smaller than the temperature of the broad component $T>$ 100 K.
Indeed, at low kinetic temperatures, the reaction (\citealt{dalgarno1984})
\begin{equation}
\label{reac1}
    {\rm H}_{3}^{+} + {\rm HD} \rightarrow {\rm H}_{2}{\rm D}^{+} + {\rm H}_{2} + 232\;{\rm K}
\end{equation}
efficiently produces H$_{2}$D$^{+}$ which can react with N$_{2}$, CN, and CO, enhancing the abundances of N$_{2}$D$^{+}$, DCN, DNC and DCO$^{+}$ to the observed values. The  D/H ratios derived from HCN and HNC in both the broad and pre-stellar components are similar. This is in good agreement with the idea that DCN and DNC not only are formed at $T<$30 K as explained above, but also through additional high-temperature routes (see e.g. \citealt{roueff2013}). Indeed, for high temperatures ($T>$70--80 K) the reaction
\begin{equation}
\label{reac2}
{\rm CH}_{3}^{+} + {\rm HD} \rightarrow {\rm CH}_{2}{\rm D}^{+} + {\rm H}_{2} + 654\;{\rm K}
\end{equation}
starts to be more efficient than reaction \eqref{reac1}. Then, CH$_{2}$D$^{+}$ reacts with atomic N and initiates a chain of reactions that ends up with high deuterium fractionation in  DCN and DNC (D/H$>$10$^{-4}$, \citealt{roueff2007}) and very low fractionation in N$_2$D$^+$ (D/H$<$3.1$\times$10$^{-5}$), as observed for the broad warm component. CH$_{2}$D$^{+}$, reacting with O and CO, could also contribute to an efficient formation of DCO$^{+}$ at high temperatures, as shown by \citet{favre2015}. However, these models show that these reactions are only efficient for the inner dense part of protoplanetary discs. Since we found for HCO$^{+}$ a very low D fractionation (D/H$<$3.1$\times$10$^{-5}$) for the broad component, this suggests that such pathways are not efficient for the typical densities found in the GC ($\sim$10$^{4}$ cm$^{-3}$).

The D/H values derived for the pre-stellar component towards G+0.693 are similar to previous observations of several deuterated complex organic molecules (COMs) towards the GC star-forming hot core Sgr B2(N2) (D/H from 5$\times$10$^{-4}$ to 4$\times$10$^{-3}$, \citealt{belloche2016}), of DCN towards the star-forming 50 km s$^{-1}$ cloud in the Sgr A$^*$ region (D/H = 4$\times$10$^{-4}$, \citealt{lubowich2000}), and of HDO towards Sgr B2 (D/H from 5$\times$10$^{-4}$ to 10$^{-3}$, \citealt{comito2003}). 

Previous measurements in the GC showed much larger D/H ratio than the primordial value, indicating a large (20$-$200) degree of chemical fractionation. 
\citet{jacq1999,lubowich2000} and \citet{polehampton2002}, claimed, through chemical modelling, that the atomic D/H ratios may be up to $\sim$100 times lower in the Galactic Centre region than in the local ISM because of stellar processing in the interior of stars. Our measured upper limit of the D/H ratio, for the first time close to the cosmic D/H ratio, is still consistent with this claim, but the large differences in the degree of fractionation in different molecules suggest that model predictions need to be considered with caution. Further observations of DCO$^+$ and N$_2$D$^+$ towards other quiescent molecular clouds in the GC, will firmly establish the degree of processing of the material in this region of the Galaxy.

Additional support for the presence of a multi-phase ISM in the GC comes from the excitation temperature derived for both components in Table \ref{table-result} which translate to very different H$_2$ densities. Using the non-LTE molecular radiative transfer model RADEX (\citealt{vandertak2007}) we derive for the warm broad component, assuming a kinetic temperature of 100 K and a $FWHM$ of 20 km s$^{-1}$, that the derived $T_{\rm ex}$ of 3--4 K translate to H$_2$ densities of 0.3--3$\times$10$^{4}$ cm$^{-3}$. For the cold pre-stellar component, assuming kinetic temperatures of 20--30 K and a $FWHM$ of 9 km s$^{-1}$, with $T_{\rm ex}$ of 7 K the required H$_2$ densities increase by at least one order of magnitude to 0.05--1$\times$10$^{6}$ cm$^{-3}$. It should be noted that the densities obtained for the narrow component are a factor from 5 up to 30 larger than for the broad component of the molecular species studied in this work. This confirms that the two components are tracing different phases of molecular gas. Moreover, higher temperatures (e.g. 100 K) and lower densities (e.g. 10$^{4}$ cm$^{-3}$), similar to those of the broad component, are not consistent neither with the D/H ratios obtained, nor with the LTE, and the non-LTE analysis performed for the narrow component, since the predicted line intensities are one order of magnitude lower than observed.
The large difference in the linewidth, the kinetic temperature and the density from the warm broad component to the pre-stellar component suggests that a substantial fraction (10\% in column density) of gas in the GC has the conditions to form the new generation of stars. This denser gas component was likely form through the cloud-cloud collisions known to  occur in the Sgr B2 complex (\citealt{fukui2021}).

To conclude, in this Letter we have shown that combined observations of the degree of deuteration of different molecules, such as N$_{2}$H$^{+}$ and HCO$^{+}$, can be used to reveal the different gas components in the line of sight of the CMZ, allowing to identify denser gas that is on the verge of gravitational collapse and which will host future protostars. A study of another similar sources in the CMZ could provide  information on the ubiquity of the multi-phase environment in the Galactic centre. Moreover, the CMZ might be used as a template for the nuclei of other galaxies since some starburst galaxies like NGC 253, M82 and IC 342 contain gas with average densities and gas temperatures similar to those found in the CMZ for the broad warm component (e.g. \citealt{aladro2011}). Thus, observations of  deuterated molecules in nearby galaxies could be crucial to identify and investigate the earliest evolutionary stages of extragalactic protoclusters, and to understand the full star-formation sequence in external galaxies.

\section*{Acknowledgments}
We are grateful to the IRAM 30m telescope staff for their help during the different observing runs, and the APEX staff for conducting the observations.
IRAM is supported by the National Institute for Universe Sciences and Astronomy/National Center for Scientific Research (France), Max Planck Society for the Advance- ment of Science (Germany), and the National Geographic Institute (IGN) (Spain).
APEX is a collaboration between the Max-Planck-Institut fuer Radioastronomie, the European Southern Observatory, and the Onsala Observatory. 
L.~C. and V.~M.R. acknowledge support from the Comunidad de Madrid through the Atracci\'on de Talento Investigador Senior Grant (COOL: Cosmic Origins Of Life; 2019-T1/TIC-15379). J.M.-P. and I.J.-S. have received partial support from the Spanish State Research Agency through project number PID2019-105552RB-C41.

\vspace{5mm}
\facilities{IRAM 30m, APEX}

\software{MADCUBA}
          
\restartappendixnumbering      
\appendix

\section{Blended transitions}
\begin{figure}
\centering
\includegraphics[width=20pc]{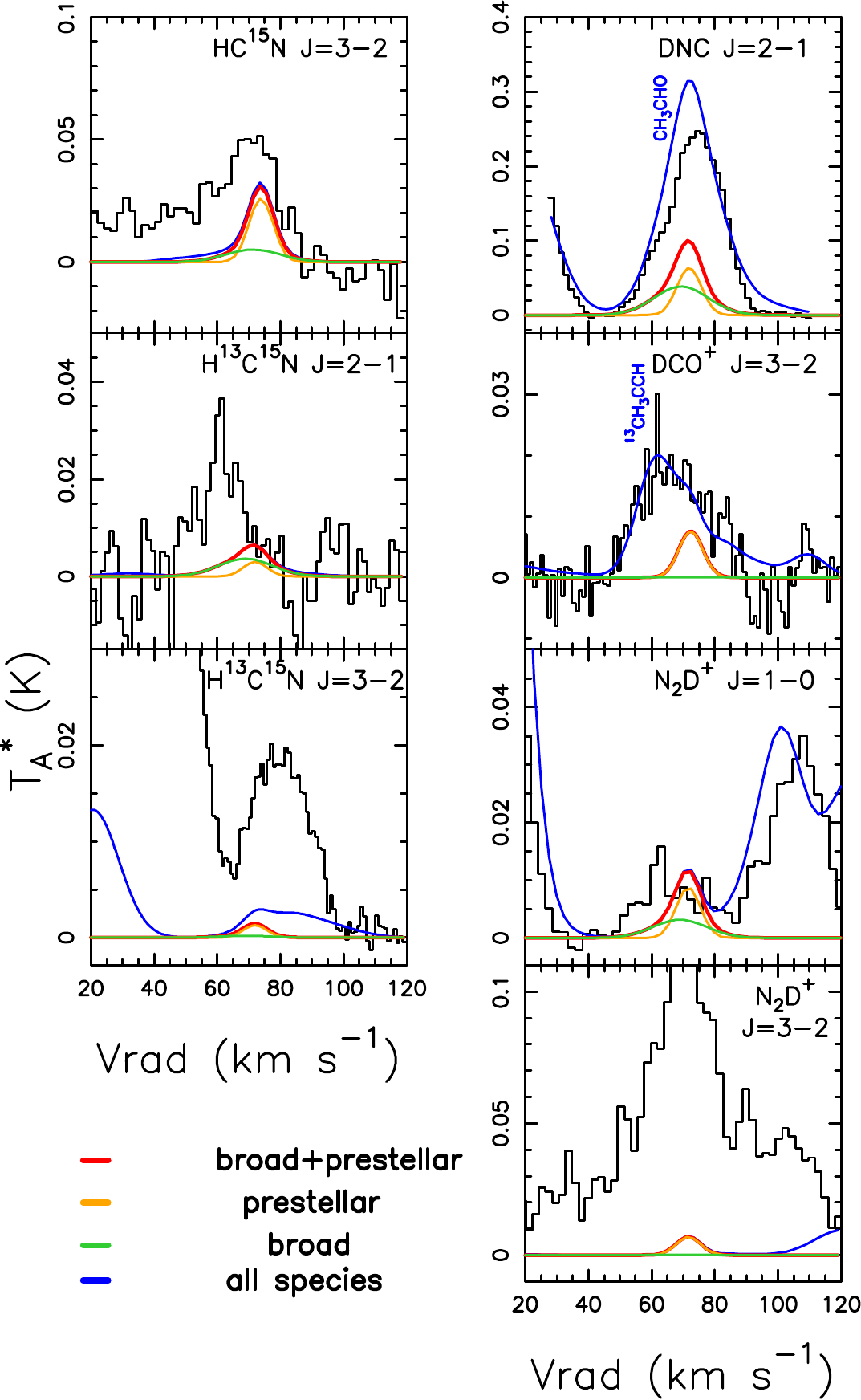}
\caption{Observed transitions that are blended with other species. The molecule and transition shown in each panel are indicated in the upper right. Blending species are indicated in blue when known. The orange line is the best LTE fit to the pre-stellar component, the green line is the best LTE fit to the broad gas component, and the red line is the sum of the components. The blue line indicates the total modelled line emission, including also the contribution of all molecular species previously identified in the survey (e.g. \citealt{rodriguez-almeida2021a, rodriguez-almeida2021b}; \citealt{rivilla2021a}; \citealt{zeng2021}).}
\label{fig-spectra-blended}
\end{figure}

Figure \ref{fig-spectra-blended} shows the transitions present in the spectral setup but not used for the analysis because they appear blended with other species.

\section{Spectroscopic information}
The transitions of the molecules studied in this work were taken from the catalogues and spectroscopic works listed in Table \ref{transitions}. Moreover, Table \ref{transitions} indicates possible blending with other species and the telescope used for the observation.

\begin{table*}
\setlength{\tabcolsep}{3.5pt}
\begin{center}
\caption{\label{transitions}Transitions of the molecules present in the G+0.693 dataset (Sect. \ref{obs}). The first column indicates the molecule for which the transitions are listed. The second and third columns show the rotational transitions and the related frequencies, respectively. log$I$ is the base 10 logarithm of the integrated intensity at 300 K and $E_{\rm up}$ is the energy of the upper level. Columns 6--8 contain the spectroscopic information for the molecules studied in this work. The last column indicate whether the transition is blended or not with other molecular species.}
\begin{tabular}{l c c  c c c c  c c  }
\hline
Molecule & Transition	& Frequency	& log$I$	& $E_{\rm up}$		&Catalogue/Entry/Date &Line list & Dipole moment & Blended	\\
& ($J'$--$J$) & (GHz)   & (nm$^{2}$ MHz) &  (K)  & &  reference & reference &  \\
\hline
HC$^{15}$N  & 1--0 &  86.0550 & -2.5525 & 4.1 &CDMS/28506/Dec. 2017 & (1), (2) &  (3) & No\\
HC$^{15}$N &  2--1 & 172.1080 &-1.6584 &12.4   & -- & -- & -- & No\\
HC$^{15}$N &  3--2 & 258.1570 &-1.1450 &24.8   & -- & -- & -- & Yes\tablenotemark{a}\\
\hline
H$^{13}$C$^{15}$N & 1--0 & 83.7276 &  -2.5875 & 4.0  & CDMS/29512/Dec. 2006 & (1) &  (3)& No\\
H$^{13}$C$^{15}$N & 2--1 & 167.4533 &	-1.6931 & 12.1 & -- & -- & -- & Yes\tablenotemark{a}\\
H$^{13}$C$^{15}$N & 3--2\tablenotemark{b} & 251.1752 &	-1.1794 & 24.1  & -- & -- & -- & Yes\tablenotemark{a}\\
\hline
DCN & 1--0 &  72.4147 & -2.7990 & 3.5  & CDMS/28509/Apr. 2006 & (4), (5) & (6)& No\\
DCN &  2--1 & 144.8280 & -1.9034 & 10.4   & -- & --  &-- & No\\
DCN &  3--2 & 217.2385 & -1.3877 & 20.9  & -- & --  &--& No\\
\hline
H$^{15}$NC & 1--0 & 88.8657 &   -2.5690 & 4.3  & JPL/28006/Dec. 1979 & (7), (8) & (9)& No\\
H$^{15}$NC & 3--2\tablenotemark{b} & 266.5878 & -1.1624 & 25.6  & --  & -- &--& No\\  
\hline
DNC & 1--0 & 76.3057 & -2.6611 & 3.7  & CDMS/28508/Sep. 2009 & (10), (11), (12) & (9)& No\\ 
DNC& 2--1 &  152.6097 &	-1.7660	& 11.0 & -- & --  &--& Yes\tablenotemark{c}\\
DNC & 3--2 & 228.9105 & -1.2510 & 22.0  & -- & --  &--& No\\
\hline
HC$^{18}$O$^{+}$ & 1--0 & 85.1622 & -2.3049 & 4.1  & CDMS/31506/Dec. 2004& (13), (14) & (15)& No\\
HC$^{18}$O$^{+}$ & 2--1 & 170.3226 & -1.4107 & 12.3 & --  & -- &--& No \\
HC$^{18}$O$^{+}$  & 3--2 & 255.4794 & -0.8972 & 24.5 & --  & -- &--& No \\
\hline
DCO$^{+}$ & 1--0 & 72.0393 & -2.5223 & 3.5 & CDMS/30510/Sep. 2009&(10), (16), (17) & (15)& No\\ 
DCO$^{+}$ & 2--1 & 144.0773 & -1.6267 & 10.4  & --  & -- &--& No\\   
DCO$^{+}$ & 3--2 &  216.1126	&-1.1110	& 20.8 & -- & --  &--& Yes\tablenotemark{d}\\  
\hline
$^{15}$NNH$^{+}$ & 1--0 & 90.2638 & -2.3485  & 4.3&CDMS/30507/Mar. 2009 & (18) & (19)& No\\
$^{15}$NNH$^{+}$ & 3--2 & 270.7836 & -0.9422 & 26.0 & --  & -- &--& No\\
\hline
N$^{15}$NH$^{+}$ & 1--0 & 91.2057 & -2.3350 & 4.4& CDMS/30508/Mar. 2009 & (18)& (19)& No\\
N$^{15}$NH$^{+}$ & 3--2 & 273.6090 & -0.9290 & 26.3 & --  & -- &--& No \\
\hline
N$_{2}$D$^{+}$ & 1--0 & 77.1092	& -2.5531 &	3.7 & CDMS/30509/Mar. 2009& (20), (21), (22) & (19)& Yes\tablenotemark{a}\\
N$_{2}$D$^{+}$ & 2--1 & 154.2170 & -1.6581 & 11.1  & --  & -- &-- & No\\
N$_{2}$D$^{+}$ & 3--2 & 231.3218 & -1.1432 &	22.2  & --  & -- &--& Yes\tablenotemark{a} \\
 \hline
  \normalsize
   \end{tabular}
   \end{center}
\tablecomments{All the transitions have been observed with the IRAM 30m, except when specified with a note.}
\tablerefs{(1) \citet{fuchs2004}; (2) \citet{cazzoli2005a}; (3) \citet{ebenstein1984}; (4) \citet{brunken2004}; (5) \citet{mollmann2002}; (6) \citet{deleon1984}; (7) \citet{creswell1976}; (8) \citet{pearson1976}; (9) \citet{blackman1976}; (10) \citet{vandertak2009}; (11) \citet{bechtel2006}; (12) \citet{okabayashi1993}; (13) \citet{plummer1983}; (14) \citet{schmid-burgk2004}; (15) \citet{botschwina1993}; (16) \citet{caselli2005}; (17) \citet{lattanzi2007}; (18) \citet{dore2009}; (19) \citet{havenith1990}; (20) \citet{pagani2009}; (21) \citet{dore2004}; (22) \citet{amano2005}.}
\tablenotetext{a}{Blended with unidentified molecular species.} \tablenotetext{b}{Transition observed with APEX.} \tablenotetext{c}{Blended  with CH$_{3}$CHO at 152.607 GHz.} \tablenotetext{d}{Blended with $^{13}$CH$_{3}$CCH at 216.115 GHz and 216.119 GHz. }
 \end{table*}

\bibliography{bibliography}{}
\bibliographystyle{aasjournal}

\end{document}